\documentclass[12pt,a4paper]{article}
\usepackage[utf8]{inputenc}
\usepackage{amsmath,graphicx,dcolumn,subcaption,mathtools}
\usepackage{braket}
\usepackage{siunitx}
\usepackage[mathscr]{eucal}
\usepackage{cite}
\usepackage{color}
\usepackage{rotating}
\usepackage[numbers,sort&compress]{natbib}
\usepackage{setspace}
\newcommand\as{\alpha_{\mathrm{S}}}

\newcommand\eps{\epsilon} 
\newcommand\ord{\mathcal{O}} 
\newcommand\f[2]{\frac{#1}{#2}}

\newcommand\bV{\mathbf{V}} 
\newcommand\bDelta{\mathbf{\Delta}} 
\newcommand\bD{\mathbf{D}} 
\newcommand\bR{\mathbf{R}} 
\newcommand\bG{\mathbf{\Gamma}} 
 
\newcommand\bF{\mathbf{F}} 
\newcommand\bJ{\mathbf{J}} 
\newcommand\bJt{\mathbf{\tilde{J}}} 
\newcommand\bI{\mathbf{I}} 
\newcommand\bb{\mathbf{b}} 
\newcommand\bTone{\mathbf{T}_1} 
\newcommand\bTtwo{\mathbf{T}_2} 
\newcommand\bTthree{\mathbf{T}_3} 
\newcommand\bTi{\mathbf{T}_i} 

\newcommand\bTonesq{\mathbf{T}_1^2} 
 
\newcommand\bqt{\mathbf{q_T}}

\newcommand\Sud{\mathcal{S}}
\newcommand\Bjet{B+\mathrm{jet}}

\def\bkp{\mathbf{k_{\bot}}}
\def\bk{\mathbf{k}}
\def\bb{\mathbf{b}}

\def\de{{\bf \Delta}}
\def\h{{\bf H}}

\def\beq{\begin{equation}} 
\def\eeq{\end{equation}} 
\def\beeq{\begin{eqnarray}} 
\def\eeeq{\end{eqnarray}}

\def\qt{$q_T$}
\def\bqt{\mathbf{q_T}}

\usepackage{hyperref}
\usepackage{footnotebackref}

\catcode`\<=\active \def<{
\fontencoding{T1}\selectfont\symbol{60}\fontencoding{\encodingdefault}}
\catcode`\>=\active \def>{
\fontencoding{T1}\selectfont\symbol{62}\fontencoding{\encodingdefault}}

\begin{document}
\begin{titlepage}
\begin{flushright}
ZU-TH 47/21\\
\end{flushright}

\renewcommand{\thefootnote}{\fnsymbol{footnote}}
\vspace*{0.5cm}

\begin{center}
  {\Large \bf Transverse-momentum resummation for\\[0.25cm] boson plus jet production at hadron colliders}
\end{center}

\par \vspace{2mm}
\begin{center}
  {\bf Luca Buonocore, Massimiliano Grazzini,\\[0.2cm] J\"urg Haag}
and    
{\bf Luca Rottoli}

\vspace{5mm}

Physik Institut, Universit\"at Z\"urich, CH-8057 Z\"urich, Switzerland

\vspace{5mm}

\end{center}

\par \vspace{2mm}
\begin{center} {\large \bf Abstract} 

\end{center}
\begin{quote}
\pretolerance 10000

We consider the associated production of a vector or Higgs boson with a jet in hadronic collisions. When the transverse momentum $q_T$ of the boson-jet system is much smaller than its invariant mass $Q$, the QCD perturbative expansion is affected by large logarithmic terms that must be resummed to all orders. We discuss the all-order resummation structure of the logarithmically enhanced contributions up to next-to-leading logarithmic accuracy. Resummation is performed at the differential level with respect to the kinematical variables of the boson-jet system. Soft-parton radiation produces azimuthal correlations that are fully accounted for in our framework.  We present explicit analytical results for the resummation coefficients up to next-to-leading order and next-to-leading logarithmic accuracy, that include the exact dependence on the jet radius.

\end{quote}

\vspace*{\fill}
\begin{flushleft}
October 2021
\end{flushleft}
\end{titlepage}

\renewcommand{\thefootnote}{\fnsymbol{footnote}}
\vspace*{2cm}

The production of vector and Higgs bosons is a crucial process at hadron colliders. It allows us to test the Standard Model, to precisely extract its parameters and is also an important background to
new physics searches. When vector or Higgs bosons are produced at high transverse momenta they are accompanied by QCD jets. This kinematical region is of particular importance for the LHC precision programme, as the reduction in event rate can be
compensated by a better identification of the boson decay products, and by an improved discrimination over the backgrounds.

Here we consider the situation in which the massive boson, or more generally, the colourless system, is produced at large transverse momentum and recoils against one or more QCD jets.
In particular, we are interested in the limit in which the total transverse momentum of the boson and the leading jet system, $q_T$, is much smaller than its invariant mass $Q$. In this region, large logarithmic contributions
due to soft and collinear radiation occur that need be resummed to all orders.
In the case of the inclusive production of a colourless system (i.e., when no additional high-$p_T$ jet is tagged) the resummation structure is fully understood \cite{Parisi:1979se,Curci:1979bg,Collins:1984kg}.
Indeed, transverse-momentum resummation for the production of colourless high-mass systems, has an all-order universal (process-independent) structure \cite{Collins:1981uk,Collins:1981va,Catani:2000vq,Catani:2013tia}.
This universality structure eventually originates from the fact that the transverse momentum of the colourless system is produced in this case just by (soft and collinear) QCD radiation
from the initial-state colliding partons.
When the colourless system is accompanied by a hard jet the resummation is significantly more complicated, due to the fact that the final state parton radiates.
The problem has been addressed by several authors \cite{Sun:2016kkh,Sun:2018icb,Chien:2019gyf,Hatta:2021jcd}.
Soft-gluon radiation is accounted for by a soft-anomalous dimension which would in principle lead to colour correlations between initial and final state partons\footnote{Since the process
features just three hard partons at Born level, the colour algebra can actually be worked out in closed form.}. Additionally, soft-parton radiation produces non-trivial azimuthal correlations \cite{Catani:2017tuc}.
The final-state collinear singularity is regulated by the finite jet radius, and the ensuing effects are described by a perturbatively computable jet function.
The situation is further complicated by the existence of the so-called Non-Global Logarithms (NGL) \cite{Dasgupta:2001sh} which enter at next-to-leading logarithmic (NLL) accuracy.
Despite these difficulties, the problem of transverse-momentum resummation for processes that feature final-state jets is theoretically interesting, and may lead to developments also for fixed-order calculations, as it happened for the production of colourless final states \cite{Catani:2007vq} and for heavy-quark production \cite{Catani:2019iny,Catani:2019hip}.

In this Letter we present new results on transverse-momentum resummation for boson plus jet production. We discuss the resummation formula that controls the logarithmically enhanced contributions up to NLL accuracy,
by accounting for the full kinematical dependence of the boson-jet system. We then present the explicit expressions of the resummation coefficients up to next-to-leading order (NLO) accuracy. 

We consider the inclusive hard-scattering process
\begin{equation}
  h_1(P_1) +  h_2(P_2) \to B(p_B) + J(p_J) + X \, ,
\end{equation}
where the collision of the two hadrons $h_1$ and $h_2$ with momenta $P_1$ and $P_2$ produces the boson $B$ with momentum $p_B$ accompanyed by a hard jet $J$ with momentum $p_J$
and $X$ denotes the accompanying final-state radiation (which may lead to additional softer jets).
Unless otherwise stated, in the following we will consider anti-$k_T$ jets~\cite{Cacciari:2008gp} and we will use the standard definition of
the distance of two particles $i$ and $j$ in rapidity $y$ and azimuth $\phi$
\begin{equation}
  \Delta R^2_{ij} = (y_{i}-y_{j})^2 + (\phi_{i}-\phi_{j})^2\, .
\end{equation}
We assume the boson $B$ to be on shell (i.e. $p_B^2=m_B^2$) and we denote with $m_J$, $y_J$ and $\phi_J$ the jet mass, rapidity and azimuthal angle.
When the invariant mass $m_J$ of the jet is integrated over,
the kinematics of the boson-jet system is specified by its total momentum $q=p_B+p_J$ and two additional independent variables that we denote by ${\bf\Omega}$.
For instance, we can use ${\bf \Omega}=\{y_J,\phi_J\}$.
The four vector $q$ is fully specified by its invariant mass $Q^2$, rapidity $y$ and transverse momentum ${\bf q_T}$.
At leading order (LO) in QCD perturbation theory the cross section is simply proportional to $\delta^{(2)}({\bf q_T})$, due to momentum conservation in the transverse plane.
At this order the process can proceed via the partonic sub-processes
\begin{align}
 q(p_1) +  \bar{q}(p_2) \to B(p_B) + g&(p_3), \quad q(\bar{q})(p_1) +  g(p_2) \to B(p_B) + q(\bar{q})(p_3), \nonumber \\
& g(p_1) +  g(p_2) \to B(p_B) + g(p_3)
\end{align}
where $p_i=z_iP_i$ and $z_i$ ($i=1,2$) are the momentum fractions carried by the incoming partons.

Beyond LO the cross section develops singular contributions as $q_T\to 0$ that need be resummed to all orders.
Up to NLL accuracy the resummation formula for the differential cross section reads  
\begin{equation}
  \label{eq:master-res}
  \begin{split} 
 \frac{d\sigma}{d^2\mathbf{q_T}d Q^2dy\,d{\bf\Omega}} &= \frac{Q^2}{2P_1\cdot P_2} \sum_{(a,c)\in\mathcal{I}}[d\sigma_{ac}^{(0)}]\int \frac{d^2\bb}{(2\pi)^2}e^{i\mathbf{b}\cdot\mathbf{q_T}}\Sud_{ac}(Q,b)\\ & \times\sum_{a_1,a_2} \int_{x_1}^1\frac{dz_1}{z_1}\int_{x_2}^1\frac{dz_2}{z_2}[\mathbf{H\Delta}C_1C_2]_{ac;a_1a_2}f_{a_1/h_1}(x_1/z_1,b_0^2/b^2)f_{a_2/h_2}(x_2/z_2,b_0^2/b^2)\mathcal{U}^{d}_{\rm NG},
  \end{split}
\end{equation}
where
$b_0=2^{-\gamma_E}$ ($\gamma_E=0.5772...$ is the Euler number) and the kinematic variables $x_1$ and $x_2$ are defined as
\begin{equation}
  x_1=\frac{Q}{\sqrt{2P_1\cdot P_2}}e^{+y}~~~~~~ x_2=\frac{Q}{\sqrt{2P_1\cdot P_2}}e^{-y}\, .
\end{equation}
The factor denoted by $\left[d\sigma_{ac}^{(0)}\right]$ is the LO cross section
\beq
\label{eq:sigLO}
\left[d\sigma_{ac}^{(0)}\right] = \as^p(Q^2) \,
\frac{d{\hat \sigma}_{ac\to \Bjet}^{(0)}(p_1,p_2;p_B,p_3)
}{Q^2 \,d{\bf\Omega}} \;\;,
\eeq
and the possible flavor combinations $(a,c)$ are $\mathcal{I} = [(q,{\bar q}),({\bar q},q),(q,g), (g,q), (g,{\bar q}), ({\bar q},g)]$ for a vector boson, $B=V$, and $\mathcal{I} = [(g,g), (q,{\bar q}),({\bar q},q),(q,g), (g,q), (g,{\bar q}), ({\bar q},g)]$ for the case of a Higgs boson $B=H$. At LO, the kinematics is given in terms of the Mandelstam invariants $s = 2 p_1\cdot p_2$, $t = -2 p_1\cdot p_3$ and $u = -2 p_2\cdot p_3$ and we can choose
${\bf\Omega}=\{t,u\}$.
In Eq.~\eqref{eq:sigLO}, $p$ refers to the appropriate power of the strong coupling: $p=1$ for $B=V$ and $p=3$ for $B=H$.

The structure of the resummation formula in Eq.~\eqref{eq:master-res} closely resembles that for the case of heavy-quark production~\cite{Catani:2014qha}, which is characterized
by a more involved colour structure given the presence of four coloured particles at Born level.
While having a simpler colour structure, the resummation formula for processes with an identified jet introduces new complications such as
a dependence on the jet definition and the presence of NGL~\cite{Dasgupta:2001sh}, which in Eq.~\eqref{eq:master-res} is encapsulated in the factor $\mathcal{U}^{d}_{\rm NG}$\footnote{ In the case of the $k_T$ \cite{Catani:1993hr,Ellis:1993tq} and Cambridge-Achen \cite{Dokshitzer:1997in,Wobisch:1998wt} algorithm additional logarithmic contributions due to the clustering are expected \cite{Ziani:2021dxr}.}.
In the following we detail all the required perturbative ingredients in Eq.~\eqref{eq:master-res}, namely the Sudakov form factor $\mathcal{S}_{ac}$,
the $[\mathbf{H\Delta}C_1C_2]$ coefficient and the NGL.

We organise the computation in such a way that the Sudakov Form factor is the same as for colour-singlet transverse momentum resummation in the formalism of Refs.\cite{Collins:1984kg,Catani:2000vq}, namely 
\begin{equation}
  \label{eq:Sudakov}
  \Sud_{ac}(Q,b) = \exp\left\{-\int_{b_0^2/b^2}^{Q^2}\frac{dq^2}{q^2}\left[A_{ac}(\as(q^2))\ln\frac{Q^2}{q^2} + B_{ac}(\as(q^2))  \right]\right\}
\end{equation}
where
\begin{equation}
  A_{ac}(\as(q^2)) = \frac{1}{2}[A_a(\as(q^2))+A_c(\as(q^2))], \quad B_{ac}(\as(q^2)) = \frac{1}{2}[B_a(\as(q^2))+B_c(\as(q^2))],
\end{equation}
and the coefficients $A_a,B_a$ coincide with the customary coefficients appearing in the case of the production of a colourless system.
  
Besides the NGL contribution which will be discussed later, the additional contributions beyond the colourless case
are embodied in the expression $\left[ \left( \h \,\de \right) C_1 C_2 \right]$, and more precisely in the factor $\de$. This
contribution starts at NLL accuracy and describes QCD radiation of
soft non-collinear partons from the underlying Born subprocess, emitted at wide angles with respect to the direction of the initial-state partons.
In the case of production of a colourless system the factor $\de$ is absent (i.e. $\de=1$).

The explicit expression of the symbolic factor 
$\left[ \left( \h \,\de \right) C_1 C_2 \right]$ is 
\begin{align}
\label{whatqq}
\left[ \left( \h \,\de \right) C_1 C_2 \right]_{c{\bar c};a_1a_2}
 = \left( \h \,\de \right)_{c{\bar c}}
\;\, C_{c \,a_1}(z_1;\as(b_0^2/b^2)) 
\;\, C_{{\bar c} \,a_2}(z_2;\as(b_0^2/b^2))
\end{align}
for the $q{\bar q}$ annihilation channel ($c=q,{\bar q}$), 
\beq
\label{whatqg}
\left[ \left( \h \,\de \right) C_1 C_2 \right]_{cg;a_1a_2}
\!= \left( \h \,\de \right)_{cg;\mu \,\nu }
\, C_{c \,a_1}(z_1;p_1,p_2,{\bf b};\as(b_0^2/b^2)) 
\, C_{g \,a_2}^{\mu \,\nu}(z_2;p_1,p_2,{\bf b};\as(b_0^2/b^2)) 
\eeq
for the quark-gluon channel ($c=q,{\bar q}$), and 
\beq
\label{whatgg}
\left[ \left( \h \,\de \right) C_1 C_2 \right]_{gg;a_1a_2}
\!= \left( \h \,\de \right)_{gg;\mu_1 \,\nu_1, \mu_2 \,\nu_2 }
\, C_{g \,a_1}^{\mu_1 \,\nu_1}(z_1;p_1,p_2,{\bf b};\as(b_0^2/b^2)) 
\, C_{g \,a_2}^{\mu_2 \,\nu_2}(z_2;p_1,p_2,{\bf b};\as(b_0^2/b^2)) 
\eeq
for the gluon fusion channel.
The functions $C_{ca}$ and $C_{ga}^{\mu \nu}$ are the universal collinear function
of colour-singlet transverse momentum resummation~\cite{Catani:2000vq,Catani:2010pd}.

The factors $\left( \h \,\de \right)$ in Eqs.~(\ref{whatqq})-(\ref{whatgg})
depend on $\bb$, $Q$ and on the kinematical variables of the underlying Born partonic process
(this dependence is not explicitly denoted in 
Eqs.~(\ref{whatqq}) and (\ref{whatgg})). Furthermore, they contain the dependence on the
jet definition.


As in Ref~\cite{Catani:2014qha}, the
shorthand notation
$\left( \h \,\de \right)$ refers to the contribution of
the factors $\h$ and $\de$, which contains
a non-trivial dependence on the colour structure of the
underlying partonic process. Explicitly, we have
\begin{align}
\label{Hqq}
\left( \h \,\de \right)_{c{\bar c}}
&=\f{\langle \,\widetilde{\cal M}_{c{\bar c}\to \Bjet} \,|
\; \de \; | \,\widetilde{\cal M}_{c{\bar c}\to \Bjet} \, \rangle
}{\as^{p}(Q^2)
\;|{\cal M}_{c{\bar c}\to \Bjet}^{(0)}(p_1, p_2;p_B,p_3)|^2}\;\; ,
\quad (c=q,{\bar q})\;\;,\\
\label{Hqg}
\left( \h \,\de \right)_{cg;\mu \,\nu }
&=\f{\langle \,\widetilde{\cal M}_{cg 
\to \Bjet}^{\nu^\prime} \,|
\; \de \; | \,\widetilde{\cal M}_{cg 
\to \Bjet}^{\mu^\prime} \, \rangle
\;d_{\mu^\prime \mu} \;d_{\nu^\prime \nu}
}{\as^{p}(Q^2)
  \;|{\cal M}_{cg\to \Bjet}^{(0)}(p_1, p_2;p_B,p_3)|^2}\;\; \quad (c=q,{\bar q})\;\;  ,\\
\label{Hgg}
\left( \h \,\de \right)_{gg;\mu_1 \,\nu_1, \mu_2 \,\nu_2 }
&=\f{\langle \,\widetilde{\cal M}_{gg 
\to \Bjet}^{\nu_1^\prime \nu_2^\prime} \,|
\; \de \; | \,\widetilde{\cal M}_{gg 
\to \Bjet}^{\mu_1^\prime \mu_2^\prime} \, \rangle
\;d_{\mu_1^\prime \mu_1} \;d_{\nu_1^\prime \nu_1}
\;d_{\mu_2^\prime \mu_2} \;
 d_{\nu_2^\prime \nu_2}
}{\as^{p}(Q^2)
\;|{\cal M}_{gg\to \Bjet}^{(0)}(p_1, p_2;p_B,p_3)|^2}\;\; 
\end{align}
where we use the colour space formalism of Ref.~\cite{Catani:1996vz}
and we denote by  $| \,{\cal M} \, \rangle$ the colour vector representing
the scattering amplitude in colour space. The `hard-virtual' amplitude 
$\widetilde{\cal M}_{cd\to \Bjet}$ is directly related to the
infrared-finite part of the all-order (virtual) scattering amplitude
${\cal M}_{cd \to \Bjet}$ of the underlying partonic process, 
and ${\cal M}_{cd \to \Bjet}^{(0)}$ is the
tree-level (LO) contribution. $|{\cal M}_{cd \to \Bjet}^{(0)}|^2$ is the squared amplitude
summed over the colours and spins of all involved partons.
The relation between ${\cal M}$ and $\widetilde{\cal M}$ is given below in 
Eq.~(\ref{eq:mtildedef}). Finally,  
$d^{\,\mu \nu}= d^{\,\mu \nu}(p_1,p_2)$ is the gluon polarization tensor, 
\begin{equation}
\label{dten}
d^{\,\mu \nu}(p_1,p_2) = - \,g^{\mu \nu} + 
\f{p_1^\mu p_2^\nu+ p_2^\mu p_1^\nu}{p_1 \cdot p_2} \;\;,
\end{equation}
which projects onto the Lorentz indices in the transverse plane.
The soft-parton factor $\de$ depends on colour matrices, and it acts as a 
colour space
operator in Eqs.~(\ref{Hqq})-(\ref{Hgg}). We note that by introducing
a colour space operator $\h$ via the definition 
$\as^p \,|{\cal M}^{(0)}|^2 \,\h = 
| \widetilde{\cal M} \, \rangle \,
\langle \,\widetilde{\cal M} |$ the shorthand
notation $\left( \h \,\de \right)$ is equivalent to
$\left( \h \,\de \right) = {\rm Tr} \left[ \h \,\de \right]$, where `${\rm Tr}$'
denotes the colour space trace of the colour operator $\h \,\de$. As in colour-singlet production,
all the process dependence is contained in the
hard factor $\h$, which is independent of the impact parameter ${\bf b}$.
The auxiliary virtual amplitude defined in Eqs.~\eqref{Hqq}-\eqref{Hgg} is related to the all-order renormalised virtual amplitude by the following factorization formula
\begin{equation}\label{eq:mtildedef}
  \ket{\widetilde{\cal M}_{cd \to \Bjet}} = \Big[1-\widetilde{\bI}_{cd\to \Bjet}\Big] \ket{{\cal M}_{cd \to \Bjet}}\, ,
\end{equation}
where the subtraction operator $\widetilde{\bI}_{cd\to \Bjet}$ can be computed order-by-order in perturbation theory
\begin{equation}
  \label{eq:isub}
  \widetilde{\bI}_{cd\to \Bjet} = \frac{\as}{2\pi}\,\widetilde{\bI}^{(1)}_{cd\to \Bjet} + \sum_{n>1} \left( \frac{\as}{2\pi} \right)^n \widetilde{\bI}^{(n)}_{cd\to \Bjet}. 
\end{equation}
The soft-parton factor $\de$ depends on the impact parameter $\bb$, 
on $Q$, on the kinematics of the Born partonic subprocess and on the jet radius $R$.
To explicitly denote the kinematical dependence, we introduce  the
azimuthal angle $\phi_{Jb}$ between the jet and the direction of $\bb$. The all-order structure of $\de$ is 
\begin{equation}
  \label{eq:delta-all-order}
\bDelta(\bb,Q;t/u,\phi_{Jb}) = \bV^\dagger(\bb,Q,t/u,R)\,\bD(\as(b_0^2/b^2),t/u,R;\phi_{Jb})\bV(\bb,Q,t/u,R).
\end{equation}
The evolution operator $\bV$, which resums logarithmic contributions $\as^n(Q^2)\ln^kQ^2b^2$ with $k\leq n$ which originate from wide-angle 
soft radiation, is obtained through the anti path-ordered exponential of the integral of the soft anomalous dimension $\bG$ 
\begin{equation}
\bV(\bb,Q,t/u,R) = \overline{P}_q \exp\left\{-\int_{b_0^2/b^2}^{Q^2} \frac{dq^2}{q^2}\bG(\as(q^2),t/u,R)\right\}.
\end{equation}
The anomalous dimension $\bG$ and the colour operator $\bD$ admit the perturbative expansions
\begin{equation}
\bG(\as,t/u,R) = \frac{\as}{\pi} \bG^{(1)}(t/u,R) + \sum_{n>1}\left(\frac{\as}{\pi}\right)^n \bG^{(n)}(t/u,R)
\end{equation}
and
\begin{equation}
  \label{eq:dexp}
  \bD(\as,t/u,R;\phi_{Jb}) = 1 + \frac{\as}{\pi} \bD^{(1)}(R;\phi_{Jb}) + \sum_{n>1}\left(\frac{\as}{\pi}\right)^n \bD^{(n)}(t/u,R;\phi_{Jb}).
\end{equation}
In particular, we observe that $\bG$, and hence $\bV$, do not depend on azimuthal angles.
In contrast, the operator $\bD$ carries the dependence on $\phi_{Jb}$ and
it is defined such that it fullfils the all-order relation $\braket{\bD} = 1$,
see Ref.~\cite{Catani:2014qha}, where $\braket{\;\cdot\;}$ denotes the azimuthal average. In particular, this implies that $\braket{\bD^{(n)}} = 0$ for $n\geq 1$.  
By performing the inverse Fourier transformation from 
$\bb$ space to $\bqt$ space, the
$\bqt$ cross section acquires an explicit dependence on $\phi_J - \phi_q$
(where $\phi_q$ is the azimuthal angle of $\bqt$).
This means that the resummation formula (\ref{eq:master-res})
leads to \qt-dependent {\it azimuthal correlations} of the produced $\Bjet$ 
system at small-\qt\, as observed in Refs.~\cite{Catani:2017tuc,Chien:2019gyf,Hatta:2021jcd}. 

The starting point for the computation of the first order resummation coefficients
is the NLO eikonal current associated to the
emission of a soft gluon off the two initial state partons, carrying momenta $p_1$ and $p_2$, and the final-state parton,
carrying momentum $p_3$,
\begin{equation}
  \label{eq:NLOSoftcurr}
  \bJ^2(\{p_i\},k;R) = \left(\bTone \cdot \bTtwo \frac{p_1 \cdot p_2}{p_1 \cdot k \;\; p_2 \cdot k} + \bTone \cdot \bTthree \frac{p_1 \cdot p_3}{p_1 \cdot k \;\; p_3 \cdot k} + \bTtwo \cdot \bTthree \frac{p_2 \cdot p_3}{p_2 \cdot k \;\; p_3 \cdot k}\right) \Theta(R_{3k}^2>R^2), 
\end{equation}
where $k$ is the momentum of the radiated gluon.
The theta function removes the contribution of soft radiation clustered within
the jet cone, thereby acting as a physical regulator of the final-state collinear singularity.
We observe that the full eikonal current in Eq.~\eqref{eq:NLOSoftcurr} includes, besides
the genuine contribution of soft radiation at wide angles we are interested in, also soft and collinear radiation from the
initial state partons. The latter contributions are already accounted for in our resummation formalism.
Therefore, we introduce a subtracted current $\bJ^2_{\rm sub }$
by removing the contribution from initial-state radiation
which, by definition, extends to the full phase space:
\begin{equation}
  \begin{split}
    \bJ^2_{\rm sub }(\{p_i\},k;R) &= \bJ^2 - \sum_{i=1,2} \left(-\bTi^2 \frac{p_1 \cdot p_2}{p_i \cdot k \;\; (p_2+p_2) \cdot k} \right) \left(\Theta(R_{3k}^2>R^2)+\Theta(R_{3k}^2<R^2)\right) \\
      & = \bigg[\bTone \cdot \bTthree \left (\frac{p_1 \cdot p_3}{p_1 \cdot k \;\; p_3 \cdot k} - \frac{p_1 \cdot p_2}{p_1 \cdot k \;\; (p_1+p_2) \cdot k}\right) \Theta(R_{3k}^2>R^2) \\ 
      & \hspace{4.5cm}+ \bTone^2 \frac{p_1 \cdot p_2}{p_1 \cdot k \;\; (p_2+p_2) \cdot k} \Theta(R_{3k}^2<R^2) \bigg ] + \bigg[ 1 \leftrightarrow 2 \bigg]  .
    \end{split}
\end{equation}
The resulting subtracted current $\bJ^2_{\rm sub }$ is regular in all collinear limits and has a simple interpretation:
it captures the soft wide angle emission between initial-final state dipole configurations. The term on the third line,
proportional to the initial state Casimir, is the leftover of the subtracted contributions in the region inside the jet cone,
and therefore far away from the initial state collinear regions where it is enhanced.
Moreover, since it does not develop a collinear singularity in the proximity of the jet direction, this contribution smoothly
vanishes when the jet radius $R$ approaches zero and hence is power suppressed in $R$.

On the other hand, the eikonal term on the second line, proportional to $\bTone \cdot \bTthree$, may develop
a final-state collinear divergence which is regularised by the jet radius, thus leading to a logarithmically enhanced
behavior in $R$. This occurs for soft configurations that are both wide-angle and collinear to the jet direction.
We can further single out these contributions from the pure soft wide-angle emission
by rewriting the subtracted current as follows
\begin{align}
  \label{eq:SsubFinal}
    \bJ^2_{\rm sub }(\{p_i\},k;R) &= \bigg[\bTone \cdot \bTthree \bigg (\frac{p_1 \cdot p_3}{p_1 \cdot k \;\; p_3 \cdot k} - \frac{p_1 \cdot p_2}{p_1 \cdot k \;\; (p_1\frac{p_2\cdot p_3}{p_1\cdot p_3}+p_2) \cdot k} + \frac{p_1 \cdot p_2}{p_1 \cdot k \;\; (p_1\frac{p_2\cdot p_3}{p_1\cdot p_3}+p_2) \cdot k}\nonumber\\
      & \hspace{2.5cm}- \frac{p_1 \cdot p_2}{p_i \cdot k \;\; (p_1+p_2) \cdot k}\bigg) \Theta(R_{3k}^2>R^2)\nonumber\\
      & \hspace{0.5cm} + \bTonesq \frac{p_1 \cdot p_2}{p_1 \cdot k \;\; (p_2+p_2) \cdot k} \Theta(R_{3k}^2<R^2) \bigg ] + \bigg[ 1 \leftrightarrow 2 \bigg]\nonumber \\
      & = \bigg[\bTone \cdot \bTthree \bigg (\frac{p_1 \cdot p_3}{p_1 \cdot k \;\; p_3 \cdot k} - \frac{p_1 \cdot p_2}{p_1 \cdot k \;\; (p_1\frac{p_2\cdot p_3}{p_1\cdot p_3}+p_2) \cdot k}\bigg) \Theta(R_{3k}^2>R^2)\nonumber \\
      & \hspace{1cm} 
      +  \bTonesq \bigg(\frac{p_1 \cdot p_2}{p_i \cdot k \;\; (p_1+p_2) \cdot k} - \frac{p_1 \cdot p_2}{p_1 \cdot k \;\; (p_1\frac{p_2\cdot p_3}{p_1\cdot p_3}+p_2) \cdot k}\bigg)\nonumber\\
      & \hspace{1cm} + \bTonesq \frac{p_1 \cdot p_2}{p_1 \cdot k \;\; (p_1\frac{p_2\cdot p_3}{p_1\cdot p_3}+p_2) \cdot k}  \Theta(R_{3k}^2<R^2) \bigg ] + \bigg[ 1 \leftrightarrow 2 \bigg]
\end{align}
where we have used colour conservation, $\sum_{i=1,3} \bTi=0$. In the last expression of Eq.~\eqref{eq:SsubFinal}, the first term is due to soft-gluon radiation collinear to the jet direction,
while the second term corresponds to soft wide-angle
initial-state radiation which is insensitive to the jet, and is integrated in the whole phase space.
The third term is a remainder in the region inside the jet cone that is power suppressed in the $R\to 0$ limit.

It follows that the resummation coefficients in $\bb$ space can be directly extracted from the following dimensionally regularised integral in
$d=4-2\epsilon$ dimensions of the subtracted current 
\begin{equation}
  \label{eq:Jtilde}
  \begin{split}
    \bJt_{\rm sub}(\bb,t/u;R) &= \mu^{2\eps} \int d^dk\delta_+(k^2)e^{i\bb\cdot\bkp}\bJ^2_{\rm sub}(\{p_i\},k;R) \\
    &= \frac{1}{4}\left(\frac{\mu^2 b^2}{4}\right)^{\epsilon}\Gamma(1-\epsilon)^2\Omega_{2-2\epsilon}\left(\frac{4}{\epsilon}\bG^{(1)}(t/u;R)-2\bR^{(1)}(\hat{\bb},t/u;R)+\dots\right),
    \end{split}
\end{equation}
where $b$ and $\hat{\bb}$ are the modulus and direction of the impact parameter vector $\bb$, respectively, and the dots stand for higher order terms in the epsilon expansion which contribute beyond the NLL level.

The calculation of the above integral is rather involved when retaining the exact dependence on the jet radius.
The result can be expressed in terms of one-fold integrals whose expressions are given in the Appendix.
In the following we present the structure of the resummation coefficients in terms of such integrals
and we report their explicit expressions in the
small-$R$ limit, which is useful for the comparison with available results in the literature.

The first order coefficient of the anomalous dimension $\bG$ reads
\begin{equation}
  \begin{split}
    \bG^{(1)}(t/u,R) &= \frac{1}{4}\left[ (\bTone\cdot\bTthree + \bTtwo\cdot\bTthree)\overline{\mathcal{A}}_{\rm out}(R) + \left( \bTone^2 -  \bTtwo^2 \right) \ln{\frac{t}{u}} -\frac{1}{2} \left( \bTone^2 + \bTtwo^2 \right) R^2 \right] \\
    &= -\frac{1}{4}\left[ (\bTone\cdot\bTthree + \bTtwo\cdot\bTthree) \ln{\frac{1}{R^2}} - \left( \bTone^2 -  \bTtwo^2 \right) \ln{\frac{t}{u}} \right] + \ord(R)\, ,
  \end{split}
  \label{eq:gamma1}
\end{equation}
where the function $\overline{\mathcal{A}}_{\rm out}(R)$ is given in Eq.~(\ref{eq:ab13ave}).
The first order coefficient $\bD^{(1)}$, which is responsible for the $q_T$-dependent azimuthal correlations at small $q_T$, is obtained as
\begin{equation}
  \bD^{(1)} = \bR^{(1)} - \braket{\bR^{(1)}}.
\end{equation}
Its explicit expression reads
\begin{align}
    \bD^{(1)}(R,\phi_{Jb})
    &= -\frac{1}{2}(\bTone\cdot\bTthree + \bTtwo\cdot\bTthree)\left[\mathcal{B}_{\rm out}(\phi_{Jb},R)-\overline{\mathcal{B}}_{\rm out}(R)\right]\nonumber\\
    & \hspace{0.5cm}- \frac{1}{2}\left(\bTone^2 + \bTtwo^2\right)R^2\left[\mathcal{B}_{\rm in}(\phi_{Jb},R)-\overline{\mathcal{B}}_{\rm in}(R)\right] \nonumber\\  
    &=  -\frac{1}{2}(\bTone\cdot\bTthree + \bTtwo\cdot\bTthree) \bigg[ -\frac{1}{2}\ln^2\left(\frac{4 \cos^2{\phi_{Jb}}}{R^2}\right) + \frac{\pi^2}{6} + \frac{1}{2}\ln^2{R^2}\nonumber\\
      & \hspace{4.5cm} +i \pi\, {\rm sign}\left(\frac{\pi}{2}-\phi_{Jb}\right)\ln\left(\frac{4 \cos^2{\phi_{Jb}}}{R^2}\right) \bigg] + \ord(R)\, ,
    \label{eq:D1}
\end{align}
where the functions ${\mathcal{B}}_{\rm in}(\phi_{Jb},R)$, ${\mathcal{B}}_{\rm out}(\phi_{Jb},R)$, $\overline{\mathcal{B}}_{\rm in}(R)$ and $\overline{\mathcal{B}}_{\rm out}(R)$
are given in Eqs.~(\ref{eq:Bin}), (\ref{eq:Bout}), (\ref{eq:ab12ave}) and (\ref{eq:ab13ave}), respectively.

Finally, we give the expression for the first order coefficient of the subtraction operator in Eq.~\eqref{eq:isub}
\begin{align}
\label{i1til}
\widetilde{\bI}^{(1)}_{cd\to 
\Bjet}\left(\eps,\frac{Q^2}{\mu_R^2};t/u,R\right) 
=  - \frac{1}{2}
\left( \frac{Q^2}{\mu_R^2}\right)^{\!\!-\eps} & \!\left\{
 \left( \frac{1}{\eps^2} +i\pi \frac{1}{\eps} 
-\frac{\pi^2}{12}\right) (\bTone^2 +\bTtwo^2)
+ \frac{2}{\eps} \,\gamma_c  
\right. \nonumber \\ &\left. 
- \frac{4}{\eps} \;\bG^{(1)}(t/u,R) + \bF^{(1)}(R) + J^{(1)}_{f}(\eps,R) \right\}, 
\end{align} 
where $f$ is the flavour (quark or gluon) of the final state jet and is
determined by the initial state partons combination $cd$.  
The additional coefficient $\bF^{(1)}$, which is an operator in colour space, originates from the subtracted soft current in
Eqs.~\eqref{eq:SsubFinal}-\eqref{eq:Jtilde} upon averaging over the azimuth, and it explicitly reads
\footnote{We note that our result differs by a constant $\pi^2/6$ term from that of Ref.~\cite{Sun:2014lna}, which is obtained by using a small virtuality to regularise the final-state collinear singularity. In our work we do not rely on this approximation, and therefore this term is absent,
in agreement with what also observed in Ref.~\cite{Liu:2020dct} in the context of deep inelastic scattering.}
\begin{align} 
  \label{F1T}
    \bF^{(1)}(R) &= -2\braket{\bR^{(1)}}(R)  = -(\bTone\cdot\bTthree + \bTtwo\cdot\bTthree) \overline{\mathcal{B}}_{\rm out}(R) - (\bTone^2 + \bTtwo^2) R^2\, \overline{\mathcal{B}}_{\rm in}(R) \nonumber\\
    &=  (\bTone\cdot\bTthree + \bTtwo\cdot\bTthree)\frac{1}{2}\ln^2{R^2}+\ord(R)\, .
\end{align}
\noindent Configurations characterised by two unresolved final state collinear partons,
which are clustered into a single jet, do not contribute to logarithmically enhanced terms
at small $q_T$, so they are not included in the singular part of the cross section that
is resummed to all-orders. On the other hand, they enter the constant term at $q_T=0$.
Their contribution to Eq.~(\ref{i1til}) is encoded in the 1-loop jet function $J^{(1)}_{f}(\eps,R)$ ($f=q,g$), which is
the same for the whole family of $k_T$ jet algorithms. We observe that a calculation of
the jet function which retains the exact dependence on the jet radius goes beyond
the collinear approximation. At the NLO level, it is certainly possible to perform a
numerical evaluation by using a suitable scheme for the subtraction of the collinear
singularities. This is beyond the scope of this work. For the sake of completeness, we
provide the expression of the 1-loop jet functions neglecting power corrections
in the jet radius $R$ \cite{Jager:2004jh,Mukherjee:2012uz}
\begin{subequations}
  \begin{align}
       \label{eq:J1g}
       J^{(1)}_g(\eps,R) &= \frac{C_A}{\eps^2} + \frac{1}{\eps}\left(2\beta_0+C_A\ln\frac{Q^2}{p_{T,J}^2R^2}\right)+\frac{1}{2} C_A \ln^2\frac{Q^2}{p_{T,J}^2R^2}+2\beta_0 \ln\frac{Q^2}{p_{T,J}^2R^2}\nonumber\\
       &+C_A\left(\frac{67}{9}-\frac{2}{3}\pi^2\right)-\frac{23}{18}n_F\\
%
       J^{(1)}_q(\eps,R) &= C_F\Bigg\{\frac{1}{\eps^2} + \frac{1}{\eps}\left(\frac{3}{2}+\ln\frac{Q^2}{p_{T,J}^2R^2}\right) +\frac{1}{2}\ln^2\frac{Q^2}{p_{T,J}^2R^2}+\frac{3}{2}\ln\frac{Q^2}{p_{T,J}^2R^2}
       +\frac{13}{2}-\frac{2}{3}\pi^2\Bigg\}\, ,
  \label{eq:J1q}
\end{align}
\end{subequations}
where $p_{T,J}$ is the jet transverse momentum (with $s p_{T,J}^2=ut$), $\beta_0 = (11C_A-2n_F)/12$ and $n_F$ is the number of active flavours.
The results in Eqs.~(\ref{eq:J1g}) and (\ref{eq:J1q}) can be easily obtained by integrating the $d$-dimensional Altarelli-Parisi splitting functions over the collinear phase space.

We note that, with the expressions in Eqs.~(\ref{eq:J1g}) and (\ref{eq:J1q}), strictly speaking, the cancellation of the poles in Eq.~\eqref{i1til} is achieved only in the small-$R$ limit.
Conversely, the computation of the exact $R$ dependence in $\bG^{(1)}(t/u,R)$ allows us to obtain the exact coefficient of the $1/\epsilon$ pole in the jet function.
We also note that the dependence on $\ln^2 R$ cancels out, as it should, in Eq.~\eqref{i1til}. 

Finally we discuss the contribution encoded in the $\mathcal{U}_{\rm NGL}^{f}$. Recently, a significant effort has been devoted to the understanding of NGL (see e.g. Refs.~\cite{Dasgupta:2012hg,Becher:2016mmh,Hatta:2020wre,Banfi:2021owj}).
In our case the NGL contribution starts to contribute at $\as^2L^2$ relatively to the Born level, and the factor $\mathcal{U}_{\rm NGL}^{f}$ can be parametrised as
\begin{equation}
\mathcal{U}_{\rm NGL}^{f}\sim \exp\Big\{-C_A C_f \lambda^2 f(\lambda,R)\Big\}~~~~~~~~~~\lambda=\frac{\as(Q^2)}{2\pi}\ln\frac{Qb}{b_0}
\end{equation}
where the function $f(\lambda,R)$ is not known in closed form and, moreover, depends on the jet algorithm\footnote{An explicit parametrisation of $\mathcal{U}_{\rm NGL}^{f}$ can be obtained by fitting results obtained with Monte Carlo simulations \cite{Dasgupta:2001sh}.}.
In the case of cone-based and anti-$k_T$ algorithms we can write \cite{Chien:2019gyf,Ziani:2021dxr}
\begin{equation}
f(\lambda,R)=f_0(R)+{\cal O}(\lambda)~~~~~~{\rm with}~~~~~~~f_0(R)=\frac{\pi^2}{3}+{\cal O}(R^2)
\end{equation}
where the coefficient $\pi^2/3$ is the same appearing in Dasgupta and Salam’s result~\cite{Dasgupta:2001sh}.
In the case of the $k_T$ \cite{Catani:1993hr,Ellis:1993tq} and Cambridge-Achen \cite{Dokshitzer:1997in,Wobisch:1998wt} algorithm the coefficient is different~\cite{Ziani:2021dxr}.

In this Letter we have discussed transverse-momentum resummation for the production of a vector or Higgs boson accompanied by a hard jet in hadronic collisions.
Barring the effects of NGL, the structure of the resummation formula can be organised in a similar way as for the case of heavy-quark production.
Collinear radiation in the jet cone produces a perturbatively computable jet function while soft wide-angle radiation leads to additional effects that can be evaluated
by integrating a suitably subtracted soft current.
We have computed the NLO resummation coefficients by keeping the exact dependence of the jet radius $R$, and retaining the azimuthal dependence.
Our results are relevant to carry out transverse-momentum resummation for a wide class of processes in which a colourless system is accompanyed by a hard jet.
Moreover, they may have an impact in extensions of the $q_T$-subtraction formalism \cite{Catani:2007vq} to this class of processes.
More details on our computation and additional results will be presented elsewhere.

\vskip 0.5cm
\noindent {\bf Acknowledgements}

\noindent We would like to thank Stefano Catani for helpful discussions and comments on the manuscript.
This work is supported in part by the Swiss National Science Foundation (SNF) under contract 200020$\_$188464.

\appendix

\section{Explicit evaluation of $\bb$-space integrals}
\label{sec:appendix}
In this Appendix we report the explicit results for the soft integrals
in $b$-space required for the calculation of the resummation coefficients
$\bG^{(1)}$ and $\bD^{(1)}$ as reported in Eqs.~(\ref{eq:gamma1}) and (\ref{eq:D1}).
We parametrize the kinematics for the generic LO process
\[ a (p_1) + c (p_2) \rightarrow B (p_B) + d(p_3), \]
in the partonic centre of mass frame as
\begin{equation}
  p_1 = \frac{\sqrt{s}}{2}(1,0,0,1)\,,\quad
  p_2 = \frac{\sqrt{s}}{2}(1,0,0,-1)\,,\quad
  p_3 = p_T(\cosh{y},1,0,\sinh{y})\,.
\end{equation}
In order to carry out the integral in Eq.~(\ref{eq:Jtilde}) there are three
independent eikonal kernels to be considered, which can be read from Eq.~(\ref{eq:SsubFinal}):
\begin{subequations}
  \begin{align}
    \label{eq:S132out}
  S_{13,2}^{\rm out}(\{p_i\},k;R) & =  \left(\frac{p_i \cdot p_3}{p_1 \cdot k p_3 \cdot k} - \frac{p_1 \cdot
  p_2}{p_1 \cdot k \left( p_1 \frac{p_2 \cdot p_3}{p_1 \cdot p_3} + p_2
    \right) \cdot k}\right) \Theta(R_{3k}>R^2),\\
  \label{eq:S123in}
  S_{12,3}^{\rm in}(\{p_i\},k;R)& =  \frac{p_1 \cdot p_2}{p_1 \cdot k \left( p_1 \frac{p_2
      \cdot p_3}{p_1 \cdot p_3} + p_2 \right) \cdot k} \Theta(R_{3k}<R^2),\\
  \label{eq:S123kin}
  S_{12,3}^{\rm kin} (\{p_i\},k) & =  \frac{p_1 \cdot p_2}{p_1 \cdot k (p_1 + p_2) \cdot k} -
  \frac{p_1 \cdot p_2}{p_1 \cdot k \left( p_1 \frac{p_2 \cdot p_3}{p_1 \cdot
  p_3} + p_2 \right) \cdot k},
\end{align}
\end{subequations}
where $k$ is the momentum of the soft gluon, which  we parametrize as
\[ k = k_T (\cosh \eta, \cos \varphi, \sin \varphi, \sinh \eta). \]
All the other kernels are obtained by applying the replacement rule $1 \leftrightarrow 2$.
The soft integrals we need to compute have the general structure 
\begin{equation}
  I_{\kappa}(b,\phi_{3b})  = \mu^{4-d}\int d^d k \delta_+(k^2)e^{i\bb\cdot\bk_\perp}S_\kappa(\{p_i\},k;R).
\end{equation}
with  $S_{\kappa}\in\{S_{13,2}^{\rm out},S_{12,3}^{\rm in},S_{12,3}^{\rm kin}\}$ and $\mu$ the mass scale introduced
by the dimensional regularisation. In general, the integrals $I_{\kappa}$
depend on the impact parameter vector $\bb$, which we parametrize in terms of its modulus $b$ and
the angle $\phi_{3b}$ that $\bb$ forms with the projection of $p_3$ onto the transverse plane. 
In the following, we first present azimutally averaged results and then those with the full
angular dependence.

\subsection{Azimutally averaged integrals}

We separate the longitudinal component of the soft momentum $k$ and express it in terms of the
rapidity difference $x=\eta-y$ 
\begin{equation}
  \label{eq:startave}
  I_{\kappa} =  \frac{\mu^{4-d}}{2}\int d^{d-2} \bk_\perp \frac{e^{i\bb\cdot\bk_\perp}}{k_\perp^2} \int_{-\infty}^{+\infty}dx (k_\perp^2S_\kappa)
            =  \frac{(\mu b)^{4-d}}{2}\int d^{d-2} \bk_\perp \frac{e^{i\hat{\bb}\cdot\bk_\perp}}{k_\perp^2} \int_{-\infty}^{+\infty}dx \hat{S}_k,
\end{equation}  
where in the last step we rescale $\bk_\perp$ by the modulus of the impact parameter $b$, and we
denote with $k_\perp^2$ the norm of $\bk_\perp$. We notice that the quantity
$\hat{S}_\kappa \equiv (k_\perp^2S_\kappa)$
is independent of $k_\perp$ while it generally
carries a dependence on the angle between $\bk_\perp$ and $p_3$.

We take the $d$-dimensional azimuthal average over the impact parameter
\begin{equation}  
  \braket{I_{\kappa}} = \frac{(\mu b)^{4-d}}{2}\int d^{d-2} \bk_\perp k_\perp^{-2} \int_{-\infty}^{+\infty}dx \hat{S}_\kappa  \braket{e^{i\hat{\bb}\cdot\bk_\perp}}
\end{equation}
where
\begin{equation}
  \braket{e^{i\hat{\bb}\cdot\bk_\perp}} = \frac{1}{B\left(\frac{1}{2},\frac{1}{2}-\epsilon\right)} \int_{0}^{\pi} d\phi \sin^{-2\epsilon}{\phi} e^{i k_\perp \cos{\phi}} = 2^{-\epsilon}\Gamma(1-\epsilon)k_\perp^\epsilon J_{-\epsilon}(k_\perp)\, .
\end{equation}
The resulting integral has a simpler structure 
\begin{equation}
  \begin{split}
    \braket{I_{\kappa}} &= (\mu b)^{2\epsilon} 2^{-1-\epsilon}\Gamma(1-\epsilon) \Omega_{1-2\epsilon}\int_0^{+\infty} d k_\perp k_\perp^{-1-\epsilon} J_{-\epsilon}({k_\perp}) \int_0^{\pi}d\varphi \sin^{-2\epsilon}\varphi \int_{-\infty}^{+\infty}dx \hat{S}_\kappa(x,\varphi;R),
  \end{split}
\end{equation}
where $\varphi$ is the angle between $\bk_\perp$ and $p_3$, and $\Omega_n=2\pi^{n/2}/\Gamma(n/2)$
is the solid angle in $n$ dimensions.
We can evaluate the integral in $k_\perp$ as the soft kernel $\hat{S}_\kappa$ does not
depend on this variable, leading to the following master formula for the averaged soft integrals
\begin{equation}
    \braket{I_{\kappa}} = \mathcal{N} \left(-\frac{1}{\epsilon}\right)B^{-1}\left(\frac{1}{2},\frac{1}{2}-\epsilon\right) \int_0^{\pi} d\varphi \sin^{-2\epsilon}\varphi\int_{-\infty}^{+\infty}dx \hat{S}_\kappa(x,\varphi;R)\\
\end{equation}
where the normalization factor is 
\begin{equation}
  \mathcal{N} = \frac{1}{4} \left(\frac{\mu b}{2} \right)^{2\epsilon} \Gamma(1-\epsilon)^2\Omega_{2-2\epsilon}.
\end{equation}

In the following, we list our results for the relevant soft kernels. For the integrals which depend on the jet radius $R$,
we give the exact result in terms of a 1-fold integral and its expansion in the small-$R$ limit. 
\begin{itemize}
\item unconstrained kernel $S_\kappa=S_{12,3}^{\rm kin}$
  \begin{align}
      \braket{I_{12}} = -\braket{I_{21}} &= \mathcal{N} \left(-\frac{1}{\epsilon}\right)B^{-1}\left(\frac{1}{2},\frac{1}{2}-\epsilon\right) \int_0^{\pi} d\varphi \sin^{-2\epsilon}\varphi\int_{-\infty}^{+\infty}dx \left(\tanh(x+y)-\tanh{x}\right)\nonumber\\
      & = \mathcal{N} \left(-\frac{1}{\epsilon}\right)2y
        = \mathcal{N} \left(\frac{1}{\epsilon}\ln\frac{t}{u} \right) ;
  \end{align}

\item inside the jet, $S_\kappa=S_{12,3}^{\rm in}$
  \begin{align}
      \braket{I_{12,3}^{\rm in}}=\braket{I_{21,3}^{\rm in}} &= \mathcal{N} \left(-\frac{1}{\epsilon}\right)B^{-1}\left(\frac{1}{2},\frac{1}{2}-\epsilon\right) \int_0^{\pi}\!\!\!\! d\varphi \sin^{-2\epsilon}\!\varphi \! \int_{-\infty}^{+\infty} \!\!\!\!\!\!\!\! dx \! \left(1+\tanh{x}\right)\Theta(R^2-x^2-\varphi^2) \nonumber\\
      & =  \mathcal{N} \left(-\frac{1}{\epsilon}\right)B^{-1}\left(\frac{1}{2},\frac{1}{2}-\epsilon\right) 2 R^2\int_0^{1} d\varphi \sin^{-2\epsilon}(R \varphi) \sqrt{1-\phi^2}\nonumber\\      
      & =  \mathcal{N} R^2 \left[  - \frac{1}{2\epsilon}  + \overline{\mathcal{B}}_{\rm in}(R)   + \mathcal{O}(\epsilon)  \right]  \nonumber\\
      &= \mathcal{N} R^2 \left[-\frac{1}{2 \epsilon} + \left( \ln{R} - \frac{1}{2} - \frac{R^2}{24} + \mathcal{O}(R^4) \right) +  \mathcal{O}(\epsilon) \right] ,      
  \end{align}
  where the function $\overline{\mathcal{B}}_{\rm in}(R)$ is given by
  \begin{equation}
    \label{eq:ab12ave}
    \overline{\mathcal{B}}_{\rm in}(R) = -\frac{1}{\pi}\int_0^{1}d\varphi 2\sqrt{1-\varphi^2} [-2\ln(2\sin{R \varphi})]\, .
  \end{equation}
 We notice that $\braket{I_{12,3}^{\rm in}}$ is a pure power correction in the jet radius $R$;
\item outside the jet, $S_\kappa=S_{13,2}^{\rm out}$
  \begin{equation}
    \begin{split}
      \braket{I_{13,2}^{\rm out}} = \braket{I_{23,1}^{\rm out}} &= \mathcal{N} \left(-\frac{1}{\epsilon}\right)B^{-1}\left(\frac{1}{2},\frac{1}{2}-\epsilon\right) \int_0^{\pi} d\varphi \sin^{-2\epsilon}\varphi \cos{\varphi}\\
      & \hspace{3cm}\times\int_{-\infty}^{+\infty}dx \frac{1+\tanh{x}}{\cosh{x}-\cos{\varphi}}\Theta(x^2+\varphi^2-R^2) \\
      & =  \mathcal{N} \left(\frac{\overline{\mathcal{A}}_{\rm out}(R)}{\epsilon} + \overline{\mathcal{B}}_{\rm out}(R)   + \mathcal{O}(\epsilon)  \right]  \\
      & = \mathcal{N} \left[  \left(\ln{R^2}-\frac{R^2}{4} + \frac{R^4}{288} + \mathcal{O}(R^6) \right)\frac{1}{\epsilon} \right.  \\
        &  \left. \hspace{3cm}+ \left(-2\ln^2{R}+\frac{R^2}{2}\ln{R} + \frac{R^2}{12} + \mathcal{O}(R^4) \right) + \mathcal{O}(\epsilon)  \right] 
    \end{split}
  \end{equation}
  where the functions $\overline{\mathcal{A}}_{\rm out}(R)$ and $\overline{\mathcal{B}}_{\rm out}(R)$ are given by the one-fold integrals 
  \begin{equation}
    \label{eq:ab13ave}
    \overline{\mathcal{A}}_{\rm out}(R) = -\frac{1}{\pi}\int_0^{\pi}d\varphi g(\varphi;R)\quad\quad \quad \overline{\mathcal{B}}_{\rm out}(R) = -\frac{1}{\pi}\int_0^{\pi}d\varphi [-2\ln(2\sin{\varphi})] g(\varphi;R)
  \end{equation}
  expressed in terms of the auxiliary function
  \begin{equation}
    \label{eq:g13}
    \begin{split}
      g_{\rm out}(\varphi,R) &= \cos{\varphi}\int_{-\infty}^{+\infty}dx \frac{1+\tanh{x}}{\cosh{x}-\cos{\varphi}}\Theta(x^2+\varphi^2-R^2) \\ 
      &= 2 \cot{\varphi} \bigg\{ \Theta(\varphi-R)(\pi-\varphi) \\
        & \hspace{2cm}+ \Theta(R-\varphi)\left(-\varphi + 2\arctan\left[\tan{\left(\frac{\varphi}{2}\right)}\coth{\left(\frac{1}{2}\sqrt{R^2-\varphi^2}\right)}\right] \right) \bigg\}\, .
    \end{split}
  \end{equation}
We note that the function $g_{\rm out}(\varphi,R)$ is the same controlling the azimuthal dependence in jet production in $ep$ scattering \cite{Hatta:2021jcd}.
\end{itemize}

\subsection{Azimuthal dependence}

We start from Eq.~\eqref{eq:startave} and observe that, in general, there are two
preferred directions in the $\bk_\perp$-space, associated to $\bb$ and the projection of
$p_3$ onto the $\bk_\perp$-space, which break the rotational invariance
of the integrand function. It is then convenient to decompose the vector
$\bk_\perp$ as
\begin{equation}
  \bk_\perp = k_1\hat{e}_1 + k_2\hat{e}_2 + k^{\perp}
\end{equation}
where $\hat{e}_{1,2}$ is an orthonormal basis in the 2-dimensional vector space 
spanned by the two preferred directions and $k^{\perp}$ lives in the remaining orthogonal
space. Choosing the vector basis such that $\hat{p_3}\cdot\bk_\perp = k_1$ and $\hat{b}\cdot\bk_\perp = k_1\cos{\phi}_{3b}+k_2\sin{\phi}_{3b}$,
we get
\begin{equation}
  \begin{split}
  I_{\kappa} &=  \frac{(\mu b)^{4-d}}{2}\int d^{d-4} \bk^{\perp} \int_{-\infty}^{+\infty}dk_1 \int_{-\infty}^{+\infty} dk_2\frac{e^{i(k_1\cos{\phi}_{3b}+k_2\sin{\phi}_{3b})}}{k_\perp^2} \int_{-\infty}^{+\infty}dx \hat{S}_\kappa\\
    & =  \frac{(\mu b)^{2\epsilon}}{2}\Omega_{-2\epsilon}\int_0^{\pi}d\varphi \sin^{-2\epsilon}\varphi\int_{-\infty}^{+\infty}\!\!\! dx \hat{S}_\kappa \int_0^{\pi}\!\!\!\! d\vartheta \sin^{-1-2\epsilon}{\vartheta}\! \int_0^\infty\!\!\!\!\! d k_{\perp} k_{\perp}^{-1-2\epsilon}e^{ik_{\perp}(\cos{\phi}_{3b}\cos\varphi+\sin{\phi}_{3b}\sin{\varphi}\cos{\vartheta})}
  \end{split}
\end{equation}
where in the last step we introduced the spherical coordinates
\begin{equation}
  k_1=k_\perp \cos{\varphi},\quad k_2=k_\perp \sin{\varphi}\cos{\vartheta},\quad k^{\perp}=k_\perp \sin{\varphi}\sin{\vartheta}.
\end{equation}
Using the following result for the integral over $k^{\perp}$ and $\vartheta$
\begin{equation}
  \begin{split}
    \int_0^{\pi}d\vartheta \sin^{-1-2\epsilon}{\vartheta} & \int_0^\infty d k_{\perp} k_{\perp}^{-1-2\epsilon}e^{ik_{\perp}(\cos{\phi}_{3b}\cos\varphi+\sin{\phi}_{3b}\sin{\varphi}\cos{\vartheta})} \\
    & = \Gamma(-2\epsilon) B(-\epsilon,-\epsilon) 2^{-1-2\epsilon} e^{-i\pi\epsilon} \cos^{\epsilon}(\phi_{3b}-\varphi)\cos^{\epsilon}(\phi_{3b}+\varphi)
  \end{split}
  \end{equation}
we obtain the master formula for the soft integrals
\begin{equation}
  I_\kappa = \mathcal{N} \left(-\frac{e^{-i\pi\epsilon}}{\pi\epsilon}\right) \int_0^{\pi}d\varphi \sin^{-2\epsilon}\varphi \cos^{\epsilon}(\phi_{3b}-\varphi)\cos^{\epsilon}(\phi_{3b}+\varphi)  \int_{-\infty}^{+\infty}dx \hat{S}_\kappa.
\end{equation}
\begin{itemize}
\item unconstrained kernel $S_\kappa=S_{12}$ 
  \begin{equation}
      I_{12} = -I_{21} =\mathcal{N}   \ln{\frac{u}{t}} \left(-\frac{e^{-i\pi\epsilon}}{\pi\epsilon}\right) \int_0^{\pi}d\varphi \sin^{-2\epsilon}\varphi \cos^{\epsilon}(\phi_{3b}-\varphi)\cos^{\epsilon}(\phi_{3b}+\varphi) = \mathcal{N} \frac{1}{\epsilon} \ln{\frac{t}{u}},      
  \end{equation}
  where we have used the result
  \begin{equation}
    \int_0^{\pi}d\varphi \sin^{-2\epsilon}\varphi \cos^{\epsilon}(\phi_{3b}-\varphi)\cos^{\epsilon}(\phi_{3b}+\varphi) = \pi e^{i\pi\epsilon} 
  \end{equation}
  Note that since the kernel does not depend on $\varphi$, one could have immediately obtained $I_{12}=-I_{21}=\braket{I_{12}}$.
\item inside the jet, $S_\kappa=S_{12}^{\rm in}$
  \begin{equation}
    \begin{split}
      I_{12}^{\rm in} = I_{21}^{\rm in} &=\mathcal{N} R^2 \left[-\frac{1}{2\epsilon}  +  \mathcal{B}_{12}(\phi_{3b};R)  +  \mathcal{O}(\epsilon) \right] \\
      & = \mathcal{N} R^2 \left[-\frac{1}{2 \epsilon} + \left( -\ln\frac{\cos\phi_{3b}}{R} - \frac{1}{2} - \ln 2 + i \frac{\pi}{2} + \mathcal{O}(R^2) \right) +  \mathcal{O}(\epsilon) \right] ,      
    \end{split}
  \end{equation}
  with
  \begin{equation}
     \mathcal{B}_{\rm in}(\phi_{3b};R) = -\frac{1}{\pi}\int_0^1 d\varphi 2\sqrt{1-\varphi^2}\big( \ln[\cos(\phi_{3b}-R \varphi)] + \ln[\cos(\phi_{3b}+R \varphi)] -2\ln[\sin(R \varphi)]- i\pi \big)
\label{eq:Bin}
  \end{equation}

\item inside the jet, $S_\kappa=S_{13,2}^{\rm out}$
  \begin{equation}
    \begin{split}
      I_{13,2}^{\rm out} = I_{23}^{\rm out} &= \mathcal{N} \left[\frac{\mathcal{A}_{\rm out}(\phi_{3b};R)}{\epsilon}  +  \mathcal{B}_{\rm out}(\phi_{3b};R)  +  \mathcal{O}(\epsilon) \right] \\
      & = \mathcal{N} \bigg\{ \left[\ln{R^2}-\frac{R^2}{4} + \frac{R^4}{288} + \mathcal{O}(R^6) \right]\frac{1}{\epsilon} \\
        & + \left[ -\frac{1}{2}\ln^2\left(\frac{4 \cos^2{\phi_{3b}}}{R^2}\right) + \frac{\pi^2}{6} +i \pi {\rm sign}\left(\frac{\pi}{2}-\phi_{3b}\right)\ln\left(\frac{4 \cos^2{\phi_{3b}}}{R^2}\right) +  \mathcal{O}(R^2) \right] +  \mathcal{O}(\epsilon) \bigg\} ,      
    \end{split}
  \end{equation}
  with
  \begin{align}
    \mathcal{A}_{\rm out}(\phi_{3b};R) &= \overline{\mathcal{A}}_{\rm out}(R) \\
    \mathcal{B}_{\rm out}(\phi_{3b};R) &= -\frac{1}{\pi}\int_0^\pi d\varphi g_{\rm out}(\varphi;R)\big( \ln[\cos(\phi_{3b}-\varphi)] + \ln[\cos(\phi_{3b}+ \varphi)] -2\ln[\sin{\varphi}]- i\pi \big)
    \label{eq:Bout}
  \end{align}
  and $\overline{\mathcal{A}}_{\rm out}(R)$ and $g_{\rm out}(\varphi;R)$ are given in Eq.~\eqref{eq:ab13ave} and Eq.~\eqref{eq:g13}, respectively.
\end{itemize}

\bibliography{biblio}

\end{document}